\documentclass[twocolumn,english,prl,aps,graphicx,multicol,epsfig]{revtex4}
\usepackage[T1]{fontenc}
\usepackage[latin1]{inputenc}
\usepackage{babel}
\usepackage{graphics}
\usepackage{color}
\usepackage{amssymb}

\begin{document}

\title{Nuclear-spin-independent short-range three-body physics in ultracold atoms}
\author{Noam Gross$^{1}$, Zav Shotan$^{1}$, Servaas Kokkelmans$^{2}$ and Lev Khaykovich$^{1}$}
\affiliation{$^{1}$Department of Physics, Bar-Ilan University,
Ramat-Gan, 52900 Israel,}\affiliation{$^{2}$Eindhoven University of
Technology, P.O. Box 513, 5600 MB Eindhoven, The Netherlands}

\begin{abstract}
We investigate three-body recombination loss across a Feshbach
resonance in a gas of ultracold $^7$Li atoms prepared in the
absolute ground state and perform a comparison with previously
reported results of a different nuclear-spin state [N.~Gross
\emph{et.al.}, Phys.~Rev.~Lett. {\bf 103} 163202, (2009)]. We extend
the previously reported universality in three-body recombination
loss across a Feshbach resonance to the absolute ground state. We
show that the positions and widths of recombination minima and
Efimov resonances are identical for both states which indicates that
the short-range physics is nuclear-spin independent.
\end{abstract}


\maketitle

A remarkable prediction of three-body theory with resonantly
enhanced two-body interactions is the existence of a universal set
of weakly bound triatomic states known as Efimov
trimers~\cite{Efimov}. In the limit of zero collision energy only
s-wave scattering is allowed signifying that a single parameter, the
scattering length $a$ is sufficient to describe the ultracold
two-body interactions. When $|a|\rightarrow\pm \infty$ the universal
long-range theory (known as "Efimov scenario") predicts that
three-body observables exhibit log-periodic behavior which depends
only on the scattering length $a$ and on a three-body parameter
which serve as boundary conditions for the short-range
physics~\cite{Braaten&Hammer06}. After decades of failed quest for a
suitable system to study the Efimov scenario~\cite{Jensen04} a
number of recent experiments with ultracold atoms have demonstrated
this logarithmic periodicity and verified the "holy grail" of the
theory, the universal scaling factor $\exp(\pi/s_{0}) \approx 22.7$
where $s_{0}=1.00624$~\cite{Zaccanti09,Pollack09,Gross09}. The
scaling as such, however, does not provide any knowledge about the
short-range part of the three-body potential which defines the
absolute location and lifetime of an Efimov state. The short range
potential is given in terms of two-body potential permutations of
the two-body subsystems and a true three-body potential which is of
importance only when three particles are very close together. In
general, it is very difficult to solve the short-range physics
exactly, and therefore this region is usually treated in terms of a
three-body parameter~\cite{Braaten&Hammer06,Dincao09}.

Among other systems of ultracold atoms which allow the study of
universal three-body
physics~\cite{Kraemer06,Ottenstein08,Zaccanti09,Huckans09,Barontini09},
bosonic lithium provides a unique opportunity to shed some light
on the short-range physics. In this Letter, we exploit the
possibility to study universality in two different nuclear-spin
states that both possess a broad Feshbach resonance.
Experimentally the three-body observable is three-body
recombination loss of atoms from a trap which is always studied in
the absolute ground state where higher order inelastic processes,
namely two-body inelastic collisions, are prohibited. However,
recently we showed that a gas of $^7$Li atoms, spin polarized in
the one but lowest zeeman sublevel ($|F=1,m_{F}=0\rangle$),
experiences very weak two-body loss which allowed a study of the
physics of three-body collisions~\cite{Gross09}. Here we
investigate three-body recombination on the absolute ground state
($|F=1,m_{F}=1\rangle$) across a Feshbach resonance at
$\sim738$~G. Comparison of the results of both states (further
denoted as $|m_F=0\rangle$ and $|m_F=1\rangle$) reveals a
remarkable identity between properties of the Efimov features. At
these large magnetic fields the two states are basically similar
in their electron spin, but different in their nuclear spin. As
the position and width of an Efimov state are solely governed by
the three-body parameter our results suggest that at high magnetic
fields the short range physics is independent on nuclear-spin
configuration and on the specific Feshbach resonance across which
universality is studied.

\begin{figure*}

{\centering \resizebox*{0.9\textwidth}{0.281\textheight}
{{\includegraphics{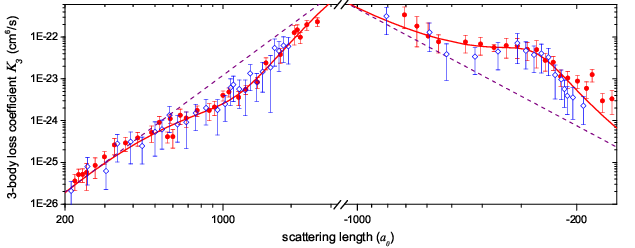}}}
\par}
\caption{\label{3bodyloss} Experimentally measured three-body loss
coefficient $K_{3}$ as a function of scattering length (in units of
Bohr radius $a_{0}$) for the $|m_{F}=1\rangle$ state (red solid
circles). The solid lines represent fits to the analytical
expressions of universal theory. The dashed lines represent the
$a^{4}$ upper (lower) limit of $K_{3}$ for $a>0$ ($a<0$). The error
bars consist of two contributions: the uncertainty in temperature
measurement which affects the estimated atom density and the fitting
error of the atom-number decay measurement. $K_{3}$ values of the
$|m_{F}=0\rangle$ state, reported by us in Ref.~\cite{Gross09}, are
represented by blue open diamonds.}
\end{figure*}

Experimentally, three-body recombination loss is studied as a
function of scattering length by means of magnetic field tuning near
a Feshbach resonance~\cite{Feshbach_review}. For positive scattering
lengths the log-periodic oscillations of the loss rate coefficient
is caused by destructive interference conditions between two
possible decay pathways for certain values of
$a$~\cite{Braaten&Hammer06,EsryNielsen99}. For negative scattering
lengths the loss rate coefficient exhibits a resonance enhancement
each time an Efimov trimer state intersects with the continuum
threshold. Recently, we found that positions of the oscillations'
minimum ($a>0$) and maximum ($a<0$) are universally related across
the Feshbach resonance on the $|m_{F}=0\rangle$ state~\cite{Gross09}
in a very good agreement with
theory~\cite{Braaten&Hammer06,Dincao09}.

Our experimental setup is described in details
elsewhere~\cite{Gross08,Gross09}. In brief, we perform evaporative
cooling at a bias magnetic field of $\sim830$~G near a Feshbach
resonance when the gas of $^7$Li atoms is spontaneously spin
purified to the $|m_{F}=0\rangle$ state~\cite{Gross08}. The atoms
are cooled down to the threshold of degeneracy and transferred
into the absolute ground state $|m_{F}=1\rangle$ by means of rapid
adiabatic passage using a radio frequency (RF) sweep scanning 1
MHz in 20 ms at a lower bias magnetic field ($35$~G). The transfer
efficiency is better than $90\%$. Finally, the bias field is
ramped to the vicinity of the absolute ground state's Feshbach
resonance ($738.3(3)$~G) where we measure atom-number decay and
temperature as a function of magnetic field from which we extract
the three-body loss coefficient $K_{3}$. Details of the
experimental procedure and the data analysis are similar to those
elaborated in Ref.~\cite{Gross09}. A typical temperature of the
atoms for positive (negative) scattering lengths is $1.8$~$\mu$K
($1.3$~$\mu$K) which matches the conditions of previously reported
measurements on the $|m_F=0\rangle$ state~\cite{Gross09}.

Experimental results of the three-body loss coefficient are
summarized in Fig.~\ref{3bodyloss} where $K_{3}$ is plotted as a
function of the scattering length $a$ for the $|m_F=1\rangle$ state
(red solid circles). Also plotted are the results of $K_{3}$
measurements for the wide resonance of the $|m_F=0\rangle$ state
(blue open diamonds) from Ref.~\cite{Gross09}. The qualitative
resemblance between the two measurements is striking. Further
investigation is achieved by treating the three-body recombination
loss as done in Ref.~\cite{Gross09}. In short, the theoretically
predicted loss rate coefficient is $K_{3}=3C_{\pm}(a)\hbar a^4/m$
where m is the atomic mass and where $\pm$ hints at the positive (+)
or negative (-) region of the scattering length. An effective field
theory provides analytic expressions for the log-periodic behavior
of $C_{\pm}(a)=C_{\pm}(22.7a)$ that we use in the form presented in
Ref.~\cite{Kraemer06} to fit our experimental data. The free
parameters are $a_{\pm}$ ($\eta_{\pm}$) which are connected to the
real (imaginary) part of the three-body parameter
\cite{Braaten&Hammer06,Marcelis08}. Moreover, $a_{-}$ defines the
position of the decay rate (Efimov) resonance and the decay
parameters $\eta_{+}$ and $\eta_{-}$, which describe the width of
the Efimov state, are assumed to be equal. Results of this fitting
procedure are summarized in Table \ref{EfimovFeatures} along with
former results of the $|m_{F}=0\rangle$ state~\cite{Gross09}.

\begin{table}[h]
\begin{center}
\begin{tabular}{|l|l|l|l|l|l|}
\hline
state & $\eta_{+}$ & $\eta_{-}$ & $a_{+}/a_{0}$ & $a_{-}/a_{0}$ & $a_{+}/|a_{-}|$\\
\hline
$|m_{F}=0\rangle$ & $0.232(55)$ & $0.236(42)$ & $243(35)$ & -$264(11)$ & $0.92(14)$ \\
\hline
$|m_{F}=1\rangle$ & $0.188(39)$ & $0.251(60)$ & $247(12)$ & -$268(12)$ & $0.92(6)$ \\
\hline
\end{tabular}
\caption{\label{EfimovFeatures}Fitting parameters to universal
theory obtained from the measured $K_{3}$ values of the
$|m_{F}=1\rangle$ and the previously reported $|m_{F}=0\rangle$
states~\cite{Gross09}.}
\end{center}
\vspace{-0.6cm}
\end{table}

The solid line in Fig.~\ref{3bodyloss} represents the fit to the
measurements performed in the $|m_{F}=1\rangle$ state. The
theoretical assumption that the real part of the three-body
parameter across a Feshbach resonance is the same for negative and
positive scattering lengths regions requires $a_{+}$ and $a_{-}$ to
obey a universal ratio
$a_{+}/|a_{-}|=0.96(3)$~\cite{Braaten&Hammer06}. Indeed, the fit
yields a remarkably close value of $0.92(6)$ which confirms the
above assumption. Moreover, the fact that $\eta_{+}$ and $\eta_{-}$
are equal within the experimental errors suggests that also the
imaginary part of the three-body parameter is identical. We thus
confirm the universality in three-body recombination across a
Feshbach resonance which we observed earlier in the
$|m_{F}=0\rangle$ state~\cite{Gross09}. Note that the recent work by
the Rice group on the $|m_{F}=1\rangle$ state~\cite{Pollack09} has
reported different values for the fitting parameters. We shall
address this apparent discrepancy later on.

Comparing the fitting parameters on different nuclear-spin states
(see Table~\ref{EfimovFeatures}) reveals striking similarities in
corresponding numbers that agree with each other exceptionally well.
We hence conclude that the three-body parameter in these states is
the same within the experimental errors. We note that both Feshbach
resonances are comparable yet slightly different in width (see
Table~\ref{FRParametersTheory}, last two rows) which has no effect
on the positions of Efimov features. Moreover, in the
$|m_{F}=0\rangle$ state there is a narrow resonance in close
proximity to the wide one (see Table~\ref{FRParametersTheory}, first
row) but it does not affect the positions of the Efimov features
either.

\begin{table}[h]
\begin{center}
\begin{tabular}{|c|c|c|c|c|}
\hline
state & type & $B_{0}$~(G) & $\Delta$~(G) & $a_{bg}/a_{0}$ \\
\hline
$|m_{F}=0\rangle$& narrow & $849.7$ & $4.616$ & $-18.94$ \\
\hline
$|m_{F}=0\rangle$& wide & $898.4$ & $-235.1$ & $-18.94$ \\
\hline
$|m_{F}=1\rangle$& wide & $742.2$ & $-169.0$ & $-20.64$ \\
\hline
\end{tabular}
\caption{\label{FRParametersTheory}Feshbach resonance parameters for
both states obtained from a fitting of the CC calculation with
Eq.~(\ref{ScatteringLength}).}
\end{center}
\vspace{-0.6cm}
\end{table}

The reported results crucially depend on precise knowledge of the
Feshbach resonance position and the scattering length in its
vicinity. We use here the same coupled-channels (CC) calculation as
in Ref.~\cite{Gross09} to predict the scattering length dependence
on magnetic field, which is then fitted with a conveniently
factorized expression~\cite{Lange09}:
\begin{equation}
\label{ScatteringLength}
\frac{a}{a_{bg}}=\prod_{i=1}^{N}\\
{\left(1-\frac{\Delta_{i}}{(B-B_{0,i})}\right)}.
\end{equation}
Here $a_{bg}$ is the background scattering length, $\Delta_{i}$ is
the $i$'s resonance width and $B_{0,i}$ is the $i$'s resonance
position. Table~\ref{FRParametersTheory} summarizes these parameters
for all Feshbach resonances in both nuclear-spin states.

\begin{figure}
{\centering \resizebox*{0.4\textwidth}{0.231\textheight}
{{\includegraphics{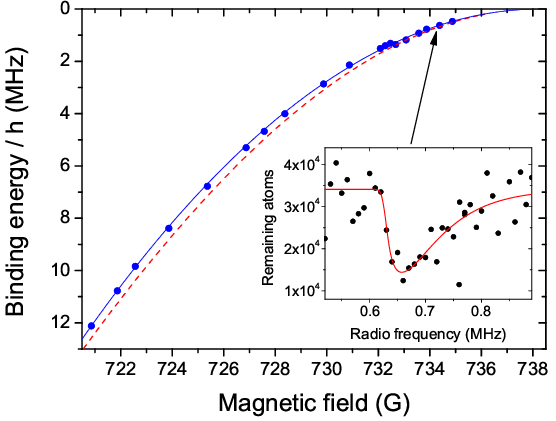}}}
\par}
\caption{\label{molecules} RF spectroscopy of the molecular
binding energy near the Feshbach resonance in the
$|m_{F}=1\rangle$ state. The solid line represents fitting to
Eq.~(\ref{BindingEnergy}). CC calculation prediction (dashed line)
is plotted as well. Inset - an example of a loss resonance at
$B=734.4$ G fitted numerically to a convolution of
Maxwell-Boltzmann and a Gaussian distributions (solid line).}
\end{figure}

To verify the $|m_{F}=1\rangle$ Feshbach resonance parameters we use
a powerful experimental technique that measures the binding energy
of the Feshbach molecules with high precision. The method uses a
weak RF field to resonantly associate weakly bound Feshbach dimers
which are then rapidly lost through collisional relaxation into
deeply bound states~\cite{Thompson05}. The remaining atom number is
measured by absorption imaging as a function of RF frequency at a
given magnetic field. In the experiment the RF modulation time is
varied between 0.5 and 3 sec and the modulation amplitude ranges
from 150 to 750~mG. RF-induced losses at a given magnetic field are
then numerically fitted to a convolution of a Maxwell-Boltzmann and
a Gaussian distributions (see inset in Fig.~\ref{molecules}). The
former accounts for broadening of the spectroscopic feature due to
finite kinetic energy of atoms at a typical temperature of $\sim
1.5\mu K$. The latter reflects broadening due to magnetic field
instability and shot-to-shot atom number fluctuations. From the fit
we extract the molecular binding energy ($E_{b}$) corresponding to
zero temperature. The RF spectroscopy of $E_{b}$ is shown in
Fig.~\ref{molecules}.

The scattering length $a$ in the vicinity of the Feshbach
resonance can be extracted from our measurement by a numerical fit
to the CC calculation. This analysis will be the subject of a
future publication. Instead we plot in Fig.~\ref{molecules}
(dashed line) the binding energies of molecules according to the
prediction of the CC calculation with no fitting parameters apart
from a shift of $-3.9$~G to the experimentally determined position
of the resonance (discussed next). A notably good agreement
between the measurements and theory indicates that very small
corrections are needed to tune the theory to the experimental
data. Here we use a simple analytical model to estimate these
corrections and to pinpoint the resonance's position.

Very close to a Feshbach resonance the molecular binding energy has
the following form \cite{Feshbach_review}:
\begin{equation}
\label{BindingEnergy}
E_{b}=\frac{\hbar^{2}}{m(a-\bar{a}+R^{*})^{2}},
\end{equation}
where $m$ is the atomic mass and $\bar{a}$ and $R^{*}$ accounts for
a finite range and a resonance strength corrections to the universal
$1/a^{2}$ law, respectively. $\bar{a}$ is the mean scattering length
which is an alternative van der Waals length
scale~\cite{gribakin93}, and
$R^{*}=\hbar^{2}/(ma_{bg}\Delta(\delta\mu))$~\cite{Petrov04}, where
$\delta\mu$ is the differential magnetic
moment~\cite{Feshbach_review}. Eq.~(\ref{BindingEnergy}) is applied
in the limit of $a\gg \bar{a}$ and $a\gg 4R^{*}$. When
Eq.~(\ref{ScatteringLength}) is substituted into
Eq.~(\ref{BindingEnergy}) the latter provides us with a fitting
expression that can be used to extract $\Delta$, $a_{bg}$ and
$B_{0}$ from our experimental data. However few comments should be
made before.

We restrict the fit to values of $E_{b}/h<4$~MHz where $a>300a_{0}$,
which corresponds to $\sim5\%$ of the Feshbach resonance width, to
meet the requirement $a\gg \bar{a} = 29.88a_{0}$ (the fitting curve
presented as a blue line in Fig.~\ref{molecules} is plotted to the
entire range of $E_{b}$). Thus, according to
Eq.~(\ref{ScatteringLength}), as $(B-B_{0})\ll |\Delta|$ the fitting
procedure is only sensitive to the product $\Gamma=a_{bg}\Delta$.
Large uncertainties are anticipated in the parameters $a_{bg}$ and
$\Delta$ if they are fitted simultaneously. We therefore arbitrary
choose to fix the values of either $a_{bg}$ or $\Delta$ to the CC
calculation prediction (see Table~\ref{FRParametersTheory}, last
row). The values in the second row of Table~\ref{FRParametersExp}
are obtained from this fitting procedure. To check the
self-consistency of the use of Eq.~(\ref{BindingEnergy}) we
calculate $R^{*}=51.8a_{0}$ using the fitting data and find it to
satisfy the second requirement of $a\gg 4R^{*}$ though less
strictly. We note that Eq.~(\ref{BindingEnergy}) was verified
against the CC calculation for $E_{b}/h<4$~MHz and they were found
to agree with each other to better than $3\%$.

Table~\ref{FRParametersExp} shows, along with the fitting data
(second row), predictions of CC calculation (first row, same as the
last row in Table~\ref{FRParametersTheory}) and experimental results
of the Rice group~\cite{Pollack09,Pollack09tunability} where $a$ was
derived from BEC \emph{in-situ} size measurements (third row).

\begin{table}[h]
\begin{center}
\begin{tabular}{|c|c|c|c|c|}
\hline
source & $B_{0}$~(G) & $\Delta$~(G) & $a_{bg}/a_{0}$  & $\Gamma$~(G) \\
\hline
CC calculation& $742.2$ & $-169.0$ & $-20.64$ & $3488$\\
\hline
RF spectroscopy& $738.3(3)$ & $-$ & $-$ & $3600(150)$\\
\hline
\emph{in-situ} BEC size& $736.97(7)$ & $-192.3(3)$ & $-24.5^{+3.0}_{-0.2}$ & $4711^{+46}_{-584}$\\
measurement \cite{Pollack09tunability}&\cite{Pollack09} & & & \\
\hline
\end{tabular}
\caption{\label{FRParametersExp}The $|m_{F}=1\rangle$ Feshbach
resonance parameters as derived from different sources.}
\end{center}
\vspace{-0.6cm}
\end{table}


The parameters of the Feshbach resonance, as determined here, appear
to be in poor agreement with the one reported by the Rice group. The
position of the resonance $B_{0}$ is shifted to a higher value of
the magnetic field beyond the experimental error~\cite{footnote}. As
for $\Gamma$, while the current measurement only slightly modifies
the CC calculation result, it differs significantly from that
reported in Ref.~\cite{Pollack09}. We note that RF spectroscopy is
very robust being independent of experimental parameters such as
trap strength, absolute number of atoms or atomic cloud size,
whereas the BEC size measurement is highly sensitive to
uncertainties in these measurements~\cite{Pollack09tunability}. The
results summarized in Table~\ref{EfimovFeatures} are obtained based
on the CC calculation and the experimentally determined value of
$B_{0}$.

The Efimov scenario was recently investigated in the $|m_F=1\rangle$
state of $^7$Li atoms by the Rice group in an impressive work
reporting in total 11 different features on both sides of the
Feshbach resonance connected to three- and four-body universal
states~\cite{Pollack09}. However, $a_{+}$ and $a_{-}$ parameters are
in apparent discrepancy with those reported here. To address this
discrepancy we compare the observed features on a magnetic field
scale instead of a scattering length scale using an inverted version
of Eq.~(\ref{ScatteringLength}) and the corresponding Feshbach
resonances' parameters indicated in Table~\ref{FRParametersExp}. For
$a>0$, the Rice and our group's loss minima are obtained at
$735.23(4)$~G and $735.39(14)$~G, respectively (we consider the
second minimum in the Rice results). There is a perfect agreement
between the two positions even within the fit errors only (see
Table~\ref{EfimovFeatures} and Ref.~\cite{Pollack09}). For $a<0$,
the Efimov resonances are located at $754.2(6)$~G and $752.4(7)$~G,
respectively, which also reasonably (within $\sim 2\times$ the error
range) agree with each other. We therefore conclude that the only
discrepancy between our groups is in the conversion of magnetic
field into scattering length, caused by the use of different
Feshbach resonance parameters. Stressing again the reliability and
precision of the method used here for resonance characterization, we
believe that a quantitative reinterpretation of the Rice group's
results will most probably resolve this discrepancy.

The nuclear-spin independent short-range physics which we report
here are partially due to the special conditions which apply already
for two-body physics at large magnetic fields. Here we are in the
Paschen-Back regime, where the electron and nuclear spins precess
independently around the magnetic field. Then the hyperfine coupling
can be neglected resulting in an uncoupled and (for $|m_F=0\rangle$
and $|m_F=1\rangle$) very similar two-body potential. This means
that non-resonant parameters, such as the background scattering
length, should be very similar for the two different states (see
Table \ref{FRParametersTheory}). However, since the derived
three-body parameters are also very similar (see Table
\ref{EfimovFeatures}), it suggests that the true three-body
forces~\cite{Dincao09} are either also nuclear-spin independent, or
they have a relatively unimportant contribution. We note that based
on similar arguments Ref.~\cite{Gemelke09} predicts a negligible
change in the three-body parameter for isolated Feshbach resonances
in Cs atoms.

We acknowledge M.~Goosen and O.~Machtey for assistance. This work
was supported, in a part, by the Israel Science Foundation and by
the Netherlands Organization for Scientific Research (NWO). N.G. is
supported by the Israel Academy of Sciences and Humanities.


\begin{thebibliography}{99}
\bibitem{footnote} We verified the position of the resonance by measuring the microwave
transition between $|F=1,m_{F}=0\rangle$ and $|F=2,m_{F}=1\rangle$
states at $B=740$~G.
\end{thebibliography}
\end{document}